\documentstyle[11pt,moriond,epsfig]{article}

\def\Journal#1#2#3#4{{#1} {\bf #2}, #3 (#4)}

\def\be{\begin{equation}}
\def\ee{\end{equation}}
\def\bea{\begin{eqnarray}}
\def\eea{\end{eqnarray}}

\def\ll#1#2{\hbox to\hsize{#1\dotfill#2}}
\def\SN{\frac{S}{N}}
\def\lNS{$\log N$--$\log S$}
\def\etal{{\it et~al.\/}{}}
\begin{document}
\vspace*{4cm}
\title{MERGING RATES OF COMPACT BINARIES IN THE
     UNIVERSE: GRAVITATIONAL WAVES AND GAMMA-RAY
     BURSTS}

\author{M.E.~PROKHOROV, V.M.~LIPUNOV, K.A.~POSTNOV}

\address{Sternberg Astronomical Institute, 119899 Moscow, Russia}

\maketitle

\abstracts{Merging rates of compact binaries
(double neutron stars or black holes) are calculated based
on the modern concept of binary stellar evolution.
It is found that the initial laser interferometers
with an rms-sensitivity of $10^{-21}$ at the frequency 100~Hz
can detect 10--700 black holes and only $\sim$1
neutron star coalescences in a 1--year integration time.
Implications of the evolutionary effects
to the cosmological origin of GRB are also discussed.}

\section{Introduction}

\footnotetext{The contribution presented by M.E.~Prokhorov}

In a few years several initial ground-based laser interferometers
aimed at searching for gravitational waves (GW) will start working
(LIGO~\cite{LIGO}, VIRGO~\cite{VIRGO}, GEO-600~\cite{GEO},
TAMA-300~\cite{TAMA}), so at present the question about
what kind of events and how frequently will
the interferometer register is very important.

Undoubtedly, the most reliable GW sources are merging
compact binary stars -- double neutron stars (NS)
and black holes (BH) of different stellar masses. In the
same time, merging compact binaries may underly the origin of
cosmic gamma-ray bursts (GRB) (Blinnikov \etal~\cite{Blin},
Pazcy\'nski~\cite{Pacz86}, Meszaros and Rees~\cite{MesRees92}).

In Section \ref{Situation} we briefly describe
observational and theoretical data about double degenerate
compact binaries and estimates of their galactic merging
rates. In Section \ref{Scenario} we discuss
evolutionary scenario  parameters that mostly affect
binary NS and BH merging rates. The results of calculations
of the merging rates in our Galaxy
are presented in Section \ref{DetectionRate}. Then we discuss
the transition from the merging rates in an individual galaxy
to the detection rate by a GW detector with characteristics
similar to those of the initial LIGO/VIRGO interferometer (this is
the most important result of the paper about GW events).
Two last parts of the paper deal with GRB (Section \ref{GRB})
and unidentified soft X-ray flashes (Section  \ref{X-ray})
recently reported in the Einstein IPC archive studies~\cite{Gotthelf}.

\section{The Present Situation
in the Galaxy}\label{Situation}

\subsection{Observational Data}

At present, the following facts are known about
double degenerate compact binaries in our Galaxy:

\begin{table}[h]
\begin{center}
\caption{Binary PSR with NS secondaries (Nice \etal~\protect\cite{5PSR})}
\label{t:psr}
\begin{tabular}{|l|ccccc|}
\noalign{\smallskip}
\hline
               &       &       &&&\\[-2mm]
\quad PSR      &$P_b(d)$& $e$  & $M_1$&$M_2$&$\tau$, yr\\[2mm]
\hline
               &       &       &&&\\[-3mm]
J1518+4904     & 8.634 & 0.249 &$\cdots$&$\cdots$&$\infty$\\
B1913+16$^a$   & 0.323 & 0.617 &1.44&1.39&$3\cdot10^8$\\
B1534+12$^a$   & 0.420 & 0.274 &1.34&1.34&$1.5\cdot10^9$\\
B2127+11c$^a$  & 0.335 & 0.681 &1.35&1.36&$3\cdot10^8$\\
B2303+46       &12.340 & 0.658 &$\cdots$&$\cdots$&$\infty$\\
B1820-11$^b$   &357.762& 0.795 &$\cdots$&$\cdots$&$\infty$\\[1mm]
\hline
\noalign{\smallskip}
\multicolumn{6}{l}{$^a$ Coalescing pulsars} \\
\multicolumn{6}{l}{$^b$ The secondary component may be not a NS} \\
\end{tabular}
\end{center}
%
%
%
\caption{BH Candidates (Cherepashchuk \protect\cite{cher}).}
\label{t:bh}
\smallskip
\centerline{%
\begin{tabular}{|l|lcccc|}
\hline
&&&&&\\[-2mm]
\multicolumn{1}{|c|}{System} & Sp. class & $P_{orb}$, d & $f_v(m), M_\odot$
    & $m_x, M_\odot$ & $m_v, M_\odot$ \\[2mm]
\hline
&&&&&\\[-3mm]
Cyg X-1     & O9,7 Iab       & 5.6 & 0.23 & 7--18  & 20--30    \\
LMC X-3     & B(3--6)II--III & 1.7 & 2.3  & 7--11  & 3--6      \\
LMC X-1     & O(7--9)III     & 4.2 & 0.14 & 4--10  & 18--25    \\
A0620-00    & K(5--7)V       & 0.3 & 3.1  & 5--17  & $\sim$0.7 \\
GS2023+338  & K0IV           & 6.5 & 6.3  & 10--15 & 0.5--1.0  \\
GSR1121-68  & K(3--5)V       & 0.4 & 3.01 & 9--16  & 0.7--0.8  \\
GS2000+25   & K(3--7)V       & 0.3 & 5.0  &5.3--8.2& $\sim$0.7 \\
GRO J0422+32& M(0--4)V       & 0.2 & 0.9  &2.5--5.0& $\sim$0.4 \\
GRO J1655-40& F5IV           & 2.6 & 3.2  & 4--6   & $\sim$2.3 \\
XN Oph 1977 & K3             & 0.7 & 4.0  & 5--7   & $\sim$0.8 \\[1mm]
\hline
&&&&&\\[-3mm]
\multicolumn{1}{|c|}{Mean BH mass}
            &                &     &      &$\sim$8.5 &  \\[1mm]
\hline
\end{tabular}}
\end{table}

\begin{enumerate}

\item A few binary radiopulsars are known to have the secondary NS
component (Table~\ref{t:psr}).

\item Three of these binary pulsars must coalesce
due to the orbital angular momentum removal by GW
on a time scale shorter than the age of the Universe (the
Hubble time $t_H\simeq15\cdot 10^9$~yr).

\item No binary pulsars with BH is known as yet (although
from evolutionary considerations one may expect one such
an object to be formed in the Galaxy per about 1000 single
pulsars, Lipunov \etal~\cite{LPPO94})

\item No binary BH has been found so far.

\item In contrast, 10 BH candidates are already known in
X-ray binary systems with normal companions~\cite{cher}. Their parameters
are listed in Table~\ref{t:bh}. Note that the mean BH mass
in these systems is
$<$$M_{bh}$$>\simeq 8.5$M$_\odot\,$, i.e. BH formed in stellar evolution
are notably more massive than NS (with the typical mass
$1.4$M$_\odot$).

\end{enumerate}

\subsection{Binary NS Merging Rate Estimates}

At present, it is possible to estimate  binary NS merging rate in two ways:
using the binary radiopulsar statistics observed and making
various computations of binary stellar evolution.

\bigskip
\centerline{``Observational'' estimates}

\ll{Phinney~\cite{Phinney}}{$1/10^6$~yr}

\ll{Narayan \etal~\cite{Narayan}}{$1/10^6$~yr}

\ll{Curran and Lorimer~\cite{CurLor}}{$3/10^6$~yr}

\ll{van den Heuvel and Lorimer~\cite{vdHLor}}{$8/10^6$~yr}

\ll{``Bailes limit''~\cite{Bailes}}{$<1/10^5$~yr}

\bigskip
\centerline{``Theoretical'' estimates}

\ll{Clark \etal~\cite{Clark}}{$1/10^4$--$1/10^6$~yr}

\ll{Lipunov \etal~\cite{LPP87}}{$1/10^4$~yr}

\ll{Hils \etal~\cite{Hils}}{$1/10^4$~yr}

\ll{Tutukov and Yungelson~\cite{TYu}}{$3/10^4$--$1/10^4$~yr}

\ll{Lipunov \etal~\cite{LNPPP95}}{$<3/10^4$~yr}

\ll{Portegies Zwart and Spreeuw~\cite{PZSp}}{$3/10^5$~yr}

\ll{Lipunov \etal~\cite{LPP96}}{$3/10^4$--$3/10^5$~yr}
\bigskip

We emphasize that although theoretical merging rates
are systematically higher than observational ones,
both estimates do not contradict each other. The main
argument is that the first (observational) estimates
of binary NS merging rate are based on the statistics of binary
systems, in which only one of the components shines as radiopulsar,
which is not at all the {\it necessary} condition for merging to occur.
Thus the observational estimates are in fact only lower limits.
There are also some selection effects that can change estimates
from both groups.

\section{Parameters of binary compact star evolution}\label{Scenario}

To calculate binary evolution, we have used
the population synthesis method (the Scenario Machine code),
which is in fact a version of the Monte-Carlo calculations.
Here we shall not enter into  detail of the evolutionary
scenario used. Much more detailed description of the method
can be found in our review~\cite{review}.

\begin{figure}[t]
\begin{center}
\epsfbox{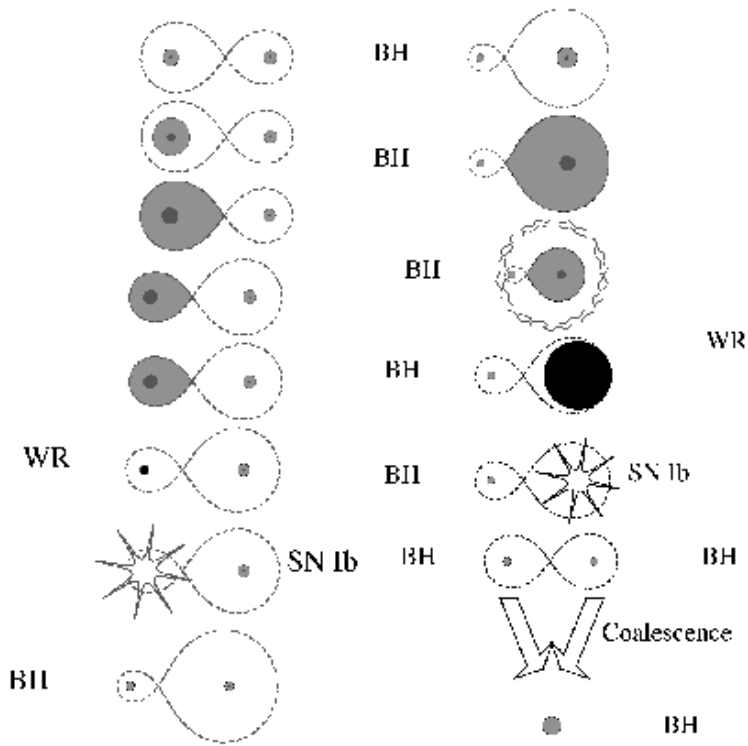}
\end{center}
\caption{Massive binary evolutionary track.}
\label{f:track}
\end{figure}

An example of the evolutionary track leading to BH+BH binary system
formation is shown in Fig. \ref{f:track}. A short glance to this track
is sufficient to understand that there are a lot of evolutionary
scenario parameters, which affect different stages of the binary
evolution. Fortunately, a very limited number of parameters has effect
on  the compact binary merging rate.

The most important (and practically unique) parameter changing the
galactic binary NS merging rate is the distribution of an additional
(kick) velocity imparted to NS at birth. The kick velocity distribution
widely accepted at present is derived from the analysis of spatial
velocities of single radiopulsars~\cite{L&L}.

We have approximated this 3-dimensional distribution 
as
$$
f_{LL}(x)dx\propto x^{0.19}(1+x^{6.72})^{-1/2}dx
$$
where $x=w/w_0$ and the characteristic velocity
$w_0$ is a parameter in our calculations. The observed
pulsar transverse velocity distribution corresponds to 
$w_0=400$~km/s.

In contrast, for BH two additional parameters appear.
First of them is a threshold main sequence stellar mass
$M_{cr}$ for the star to collapse into a BH after
its nuclear evolution has ended. This parameter is still poorly
determined and varies in a wide range:
e.g., according to van den Heuvel and Habets
\cite{vdH&Habets} $M_{cr}= 40$--80M$_\odot$;
Tsujimoto \etal~\cite{Nomoto} give 40--60M$_\odot$; Portegies Zwart
\etal~\cite{PZwetal} derive $>$20M$_\odot$.

The second parameter is the fraction of the presupernova mass,
$k_{BH}$, collapsing into BH. This parameter is fully unknown, so
we varied it from 0.1 to 1 in our calculations.

The third parameter, as for NS, is the kick velocity. Clearly,
in the general case the more massive BH will acquire smaller
velocities than NS. In our calculation we used the following
{\it ad hoc} relationship
$$
    \frac{w_{BH}}{w_{NS}} =
        \frac{M_{preSN}-M_{BH}}{M_{preSN}-M_{OV}}\,,
$$
where $M_{OV}=2.5$M$_\odot$ is the maximal NS mass
(Oppenheimer-Volkoff limit). When BH mass is close to
$M_{OV}$, velocities of BH and NS are assumed to be almost the same,
whereas at $k_{BH}=1$  BH kick velocity is assumed to vanish.
(Of course, other dependences $w_{BH}/w_{NS}$ are possible, but
their specific shape weakly affects the results).


\section{Detection rate of binary compact star merging}\label{DetectionRate}

\begin{figure}
\begin{center}
\epsfxsize=0.5\hsize
{\epsfbox{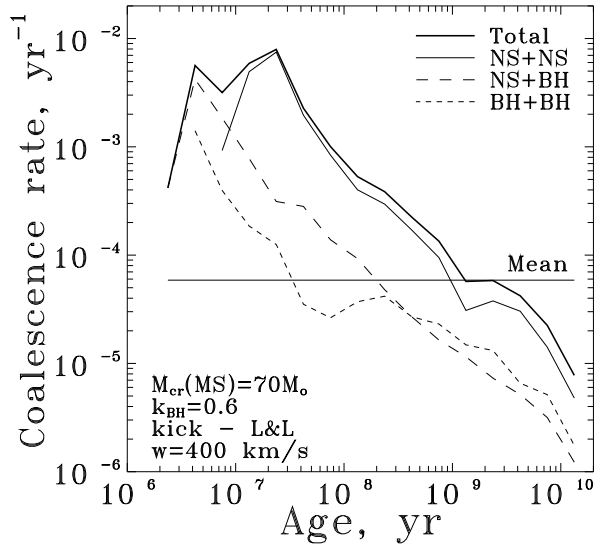}}
\end{center}
\caption{The dependence of different compact binary
merging rates in a model ``elliptical'' galaxy
(where all the stars were formed simultaneously
at the moment $t=0$) with mass $M=10^{11}$M$_\odot$.
The horizontal line shows the mean merging rate of binary NS
in a spiral galaxy of the same mass with a stationary star formation.
}\label{f:green}
\end{figure}

\begin{figure}
\begin{center}
\epsfxsize=0.5\hsize
{\epsfbox{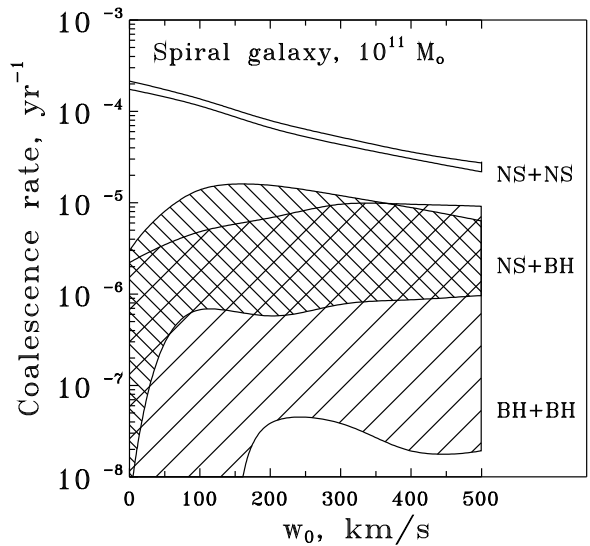}}
\end{center}
\caption{The dependence of different compact binary merging rates
in a spiral galaxy with $10^{11}$M$_\odot$ on the characteristic
kick velocity $w_0$.
}\label{f:R(w)}
\end{figure}

Under the assumptions made above, we can calculate
the binary merging rates in the Galaxy $R$. The results are
presented in Fig.~\ref{f:green} and~\ref{f:R(w)}

After having found the merging rates $R$ in a typical galaxy,
we need to go over the event rate $D$ at the detector.
Applying the optimal filtering technique~\cite{Thorne87}, the
signal-to-noise ratio $S/N$ at the spiral-in stage is
$$
    \SN \propto \frac{M_{ำh}^{5/6}}{d} \,.
$$
Here $M_{ch}= (M_1M_2)^{3/5}(M_1+M_2)^{2/5}$ is ``chorp''--mass of the
binary system.
This means that for a given $S/N$ our detector
can register more massive BH from larger distances
than NS. The volume within which BH or NS is to be detected
should be proportional $M_{ch}^{15/6}\,$. Then the ratio
of detection rates of BH and NS can be written as
$$
    \frac{D_{BH}}{D_{NS}} = \frac{R_{BH}}{R_{NS}}
           \left( \frac{M_{BH}}{M_{NS}} \right)^{15/6}\,.
$$

\begin{figure}[t]
\begin{center}
\epsfxsize=0.5\hsize
{\epsfbox{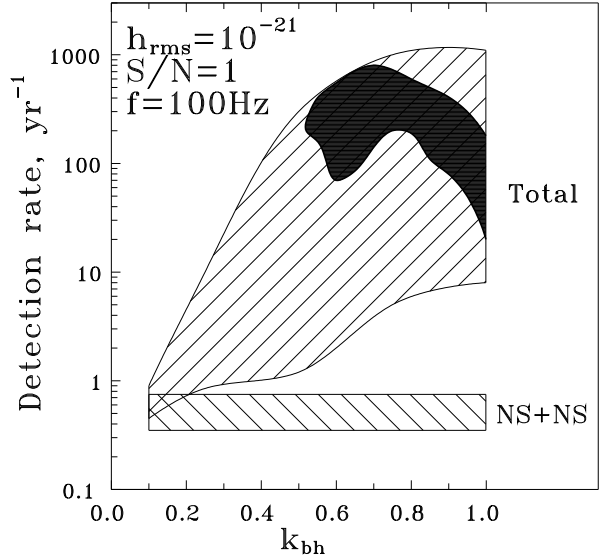}}
\end{center}
\caption{The total merging rate of NS+NS, NS+BH, and BH+BH binaries as
would be detected by a laser interferometer with $h_{rms}=10^{-21}$ for
Lyne--Lorimer kick velocity distribution with $w_0=200$--400~km/s and BH
progenitor's masses $M_*=15$--50M$_\odot$, for different scenarios of
binary star evolution as a function of $k_{bh}$. NS+NS mergings are
shown separately.  In all cases BH+BH mergings contribute more than
$80\%$ to the total rate. The filled ``Loch--Ness--monster--head''--like
region corresponds to BH formation parameters
$M_*>18$M$_\odot$ and $k_{bh}=0.5$.}
\label{f:sheja}
\end{figure}

Let us make a simple estimate. Take a GW detector
with $h_{rms}=10^{-21}$ and $S/N=1$ (as for the initial  LIGO or
VIRGO interferometer). Let the mass of NS and BH be
1.4M$_\odot$ (the typical value well justified experimentally) and
8.5M$_\odot$ (the mean mass of the BH candidates,
see Table~\ref{t:bh}). Let us also assume that any star with
the initial mass $M(NS)>10$M$_\odot$ yields NS (the typical value
confirmed theoretically), and the threshold mass for BH formation
is the maximal from the estimates given above,
$M(BH)>M_{cr}=80$M$_\odot\,$. Hence using Salpeter mass function
we find
$$
    \frac{R_{BH}}{R_{NS}} \simeq \frac{N(M>80M_\odot)}{N(M>10M_\odot)}
    = \left(\frac{80M_\odot}{10M_\odot}\right)^{-1.35} \simeq 0.06
$$
and then
$$
   \frac{D_{BH}}{D_{NS}} =
    \left(\frac{80M_\odot}{10M_\odot}\right)^{-1.35}
    \left(\frac{8.5M_\odot}{1.40M_\odot}\right)^{15/6}
    \simeq 0.06 \times 90 \simeq 5
$$

This estimate is, of course, very rough, but the precise calculation
gives essentially the same result.  Fig.~\ref{f:sheja} shows the
calculated absolute registration rate of mergings $D$ by the detector
with $h_{rms}=10^{-21}$ and $S/N=1$, as a function of  $k_{BH}$.  The
calculations have been done for Lyne--Lorimer kick velocity distribution
with the parameter $w_0=400$~หอ/ำ.  The vertical dispersion is due to
the dependence of $D$ on $M_{cr}$.  It is seen that binary NS mergings
occur with a rate of about 1 event per year.  In the same time, the
total detection rate (except for $k_{BH}\approx0.1$) can exceed that of
binary NS mergings by 2--3 orders.  The filled region shows the region
of the ``most probable parameters'', for which the calculations meet both
the upper limit to the number of BH with radiopulsars (less than 1 per
700 single pulsars) and the number of BH candidates of Cyg~X-1 type
(from  1 to 10 in the Galaxy).  In this region, from 10 to 100 mergings
per year must be detected, with the majority of them being binary BH
mergings.

\subsection{Discussion}

\begin{figure}
\begin{center}
\epsfxsize=\hsize
{\epsfbox{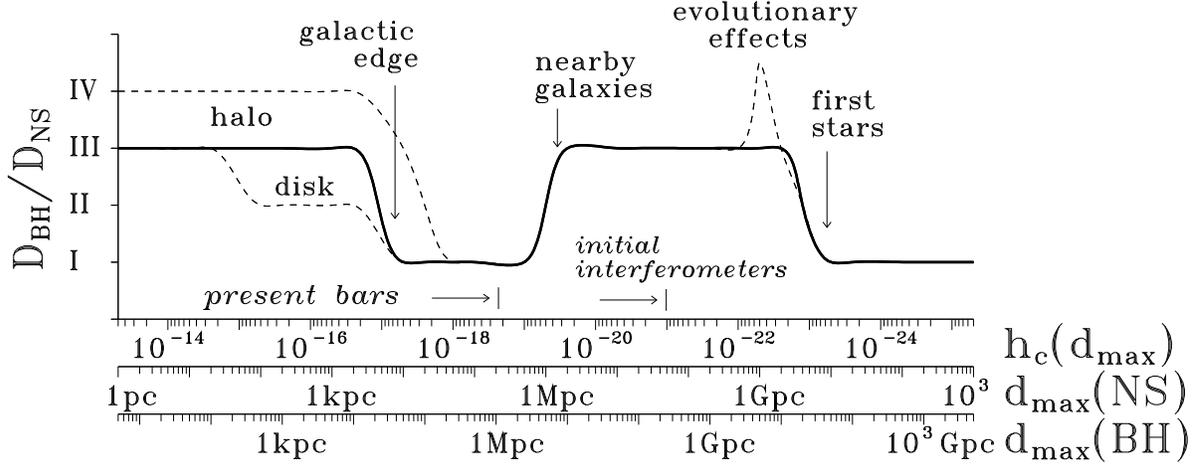}}
\end{center}
\caption{A schematic dependence of the
detection rate ratio on the detector sensitivity  (the upper
scale) and the limiting distance from which binary NS ($1.4+1.4$M$_\odot$,
the middle scale) and  binary BH ($8.5+8.5$M$_\odot$, the bottom scale)
mergings can be detected. See the text for more detail.}
\label{f:Dbh/Dns}
\end{figure}

Nevertheless, one may wonder whether binary BH (which are less numerous
than binary NS) should always be detected more frequently than binary NS.
It seems worth looking more closely on the stellar mass distribution
around a GW detector on Earth (Fig.~\ref{f:Dbh/Dns}).
In this Figure, the detection rate ratio $D_{BH}/D_{NS}$ is plotted
schematically
against the detector sensitivity level (the upper scale), which
can be expressed through the maximum distance from which
binary NS/BH mergings with $M_{NS}=1.4$M$_\odot$ and
$M_{BH}=8.5$M$_\odot$  can be detected (two bottom
scales). Four segments may be distinguished in this plot (from left
to right): first, when we register objects inside some part of the Galaxy,
second, when all objects within the Galaxy are detected but no
extragalactic objects can be detected, third, mainly extragalactic
events are detected from distances more close than those at which 
the initial star
formation occurs, and forth, where we detect all events in the Universe.
In different segments different detection ratios will be obtained.
In the first segment, the detection ratio depends on the galactic
structure: if NS and BH populate the same spherical halo, this ratio
(roman III on the vertical axis; the solid line) is
$
  {R_{BH}}/{R_{NS}} ( {M_{BH}}/{M_{NS}} )^{15/6}\,;
$
if NS and BH populate the galactic disk, this ratio becomes
(roman II, the bottom dashed line)
$
  {R_{BH}}/{R_{NS}} ( {M_{BH}}/{M_{NS}} )^{10/6}\,;
$
if NS fill more extended halo than BH, i.e. the halo radius $r_{NS}>r_{BH}$
(roman IV, the upper dashed line), then the detection ratio is
$
  {R_{BH}}/{R_{NS}} ( {M_{BH}}/{M_{NS}} )^{15/6}
  (r_{BS}/r_{NS})^3\,.
$
In the second and fourth segments the detection ratio
will be minimal and simply equal to (roman I)
$
  {R_{BH}}/{R_{NS}}\,.
$
In the third segment, the detection ratio of type III  is
realized. At the end of this segment evolutionary effects
can affect the detection ratio (the dashed line).

Note that the present  sensitivity of bar detectors ($h_{rms}\sim
10^{-19}$) falls within the second segment (an ``unhappy'' situation
because no mass-ratio enhancement for BH detection occurs), whereas the
initial laser interferometers ($h_{rms}\sim 10^{-21}$) are ``luckily''
in the third segment with the enhanced type III detection ratio.  With the
advanced LIGO sensitivity, it is possible to detect evolutionary
effects and even reach the very edge of star formation.

\section{Gamma-Ray Bursts}\label{GRB}

\begin{figure}
\begin{center}
\vbox{\hbox to\hsize{%
\hss
\epsfxsize=0.5\hsize
\epsfbox{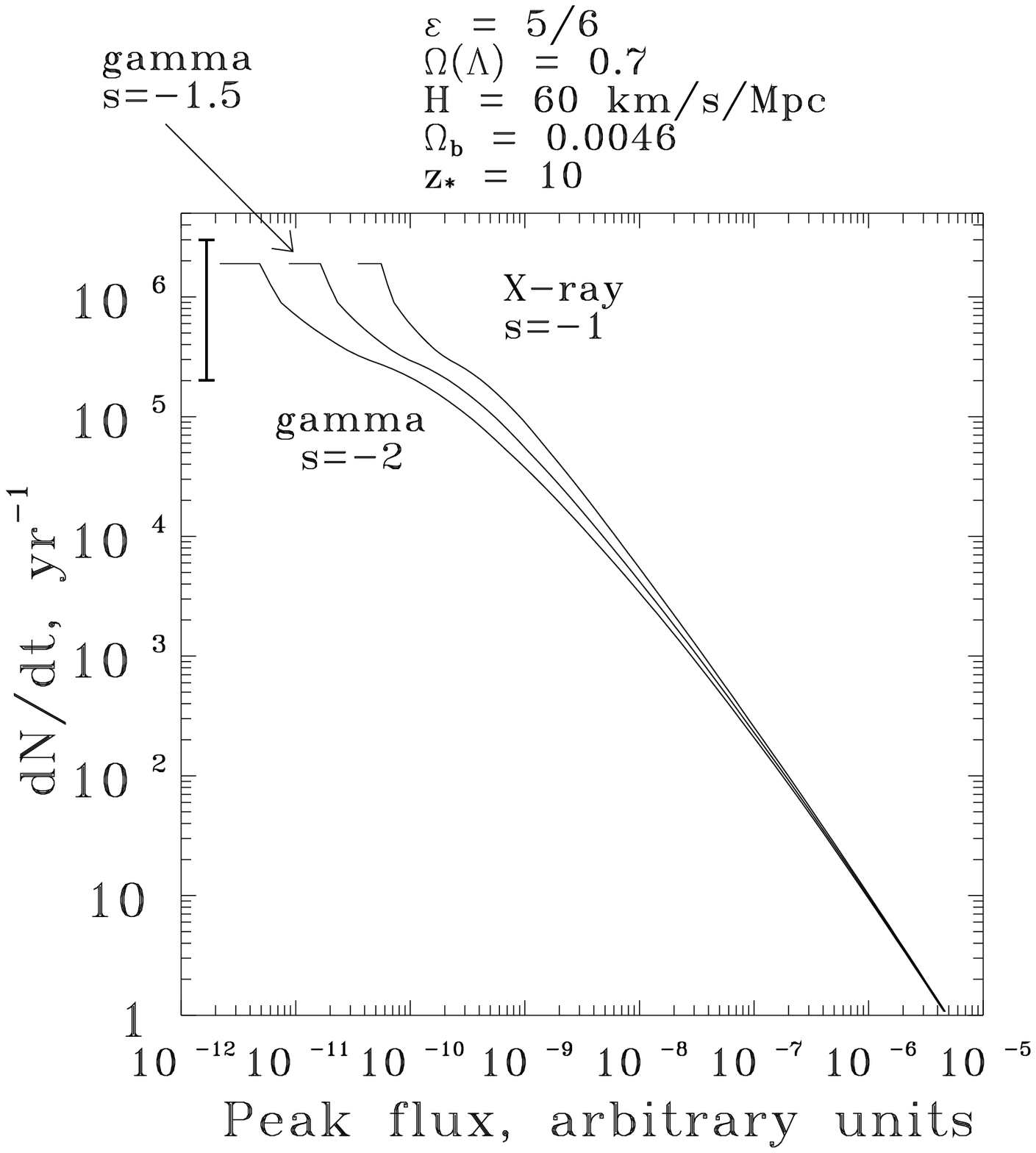}
\hss
\epsfxsize=0.5\hsize
\epsfbox{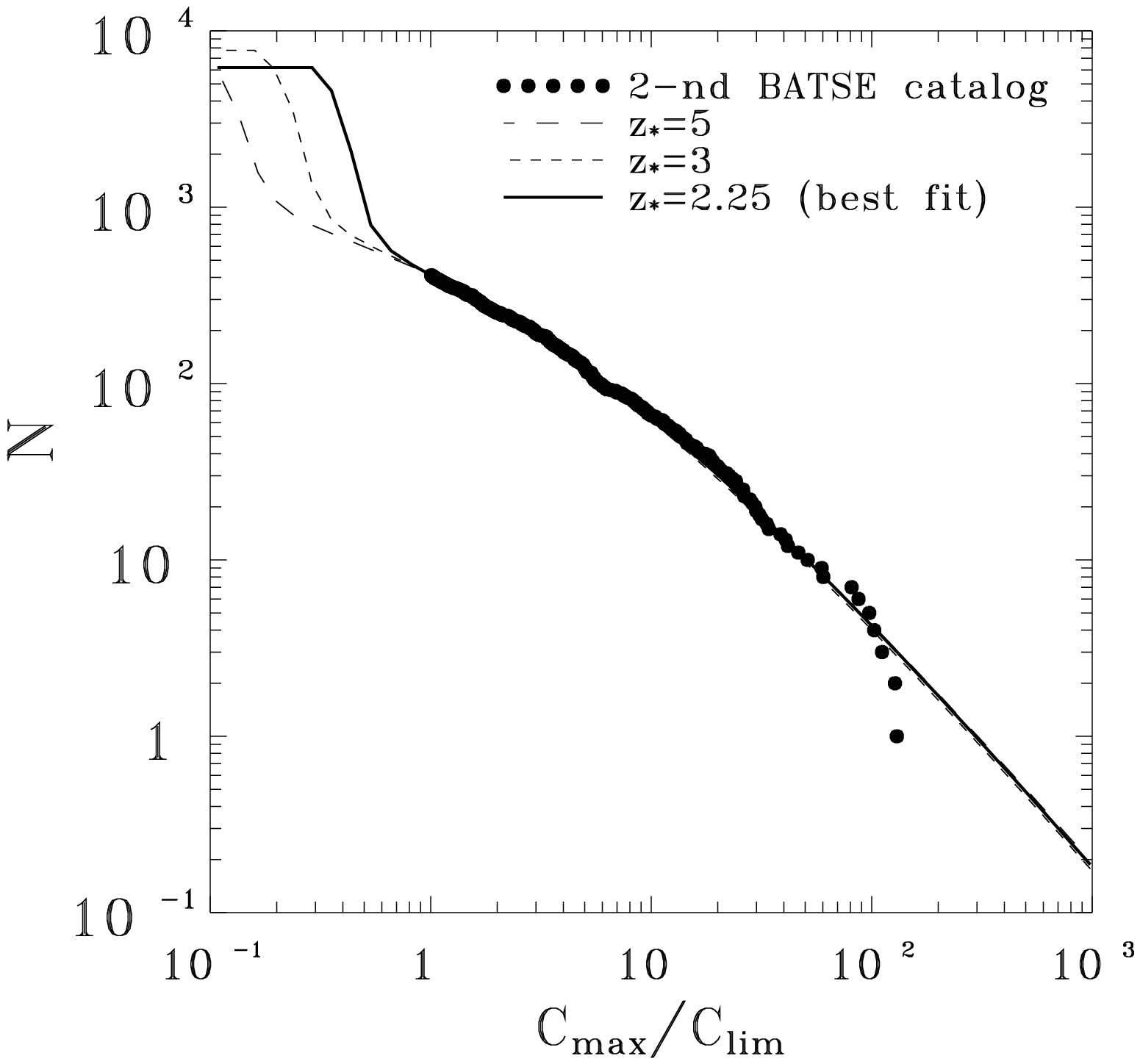}
\hss
}}
\end{center}
\caption{(Left panel) \lNS{} curves calculated for different
spectral power-law indices attributable to gamma- and X-ray emission
in a GRB. Values of the cosmological model parameters are
shown in the Figure.
The bar indicates the accessible range of the total GRB rates
in the entire Universe varying the parameters as discussed in the text.}
\label{f:logNlogS}
\caption{(Right panel) The 2-nd BATSE catalog (solid points) is fitted by
the cosmological GRB model (from Lipunov 
\etal~\protect\cite{LPP1995}). Note that the total GRB rate
in the Universe is $\sim$$10^4$ per year, $\sim$3 orders of magnitude
smaller than the total binary NS merging rate.}
\label{f:BATSE}
\end{figure}

Using the obtained above dependence of compact binary merging rates
in the elliptical galaxy on time (Fig.~\ref{f:green})
and assuming the cosmological origin of GRB as products
of binary NS/BH coalescences, we can compute the theoretical
\lNS\ curve. To do this, we need to specify the cosmological
parameters, the moment of the star formation beginning,
and the spectral power-law index of a typical GRB
(see Lipunov \etal~\cite{LPP1995} for more detail).
Taking the density of baryons in stars
$\Omega_*=0.0046$ (in terms of critical density to close
the Universe)~\cite{Fukugita} and varying other parameters
within limits permitted by the present theory and observations
($\Lambda$--term: $0\le\Omega_\Lambda<0.7$;
the fraction of elliptical galaxies: $0.15<\varepsilon<0.9$;
the star formation starting redshift:
$2.5<z_*<10$), the total compact binary merging rate
in the Universe (a constant the \lNS\ curve goes at small fluxes)
is found to vary not too much from
a few$\cdot10^5$~yr$^{-1}$ to a few$\cdot10^6$~yr$^{-1}$ (see
Fig.~\ref{f:logNlogS}).

This curve is consistent with observational data obtained by BATSE
(Fig.~\ref{f:BATSE} from~\cite{LPP1995}). However, the total
GRB rate in the Universe will be of order
$\sim$$10^4$~yr$^{-1}$. The difference in about 3 orders
can be explained either by assuming the only one merging of $\sim$1000
to yield GRB, or by the gamma-ray emission collimation into a
$\sim$$7^\circ$ solid angle~\cite{LPP1995}.

\section{Einstein Soft X-ray Bursts}\label{X-ray}

\begin{figure}
\begin{center}
\epsfxsize=0.7\hsize
{\epsfbox{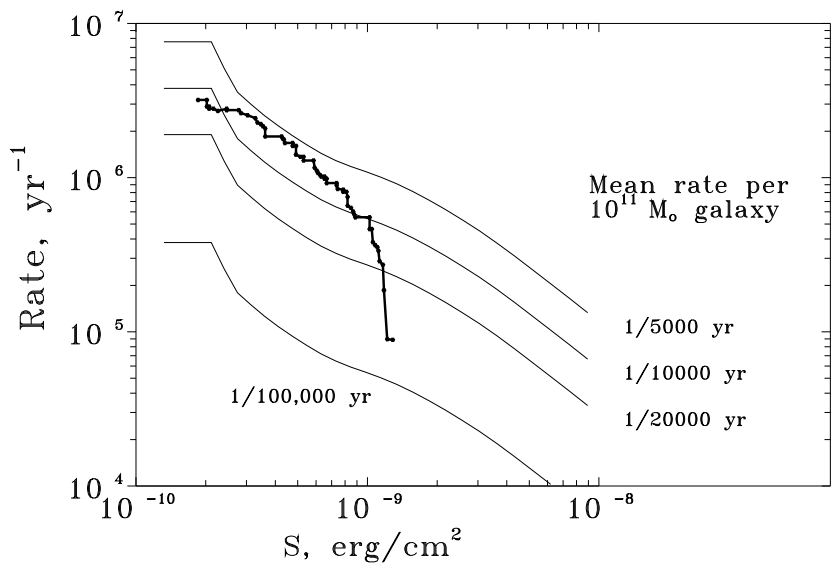}}
\end{center}
\caption{The observed \lNS\ distribution for the
Einstein IPC soft X-ray flashes (from Gotthelf \etal~\protect\cite{Gotthelf};
the broken line) and theoretical \lNS\ curves (with the same parameters
as in Fig.~\protect\ref{f:logNlogS}) for four different values
of the mean binary NS merging rate in a spiral
$10^{11}$M$_\odot$ galaxy.}
\label{f:x-ray}
\end{figure}

In 1996, after Einstein IPC archive data reprocessing,
weak soft X-ray flashes were reported~\cite{Gotthelf}, 
which are distributed isotropically on the sky and
unidentified with known astronomical objects, like GRB.
The interpolation of the Einstein IPC field of view
on the total celestial sphere yields
the rate of these flashes $\sim 2\cdot10^6$~yr$^{-1}$.
If the existence of these flashes is confirmed
(for example, using independent ROSAT and ASCA data) and they are
generated by binary NS coalescence, they will be a serious
indication favouring the high, $\sim$1/10000~yr$^{-1}$,
NS merging rate in the Galaxy (see Fig.~\ref{f:x-ray}).

\section{Conclusion}

To conclude, we expect the GW events due to compact binary coalescences
to be registered already by the initial laser GW-interferometers of
LIGO/VIRGO type at a rate substantially higher than has previously been
thought. We also think that BH will thus be detected simultaneously
with GW. Otherwise, we will have very stringent constraints on BH
formation parameters, which is, of course, much less interesting.

\section*{Acknowlegements}
MEP thanks the Organizing Committee of the XXXII Moriond Conference
for financial support. The work was partially supported by the grant of Russian
Fund for Basic Research No 95-02-06053a.

\end{document}